\documentclass[prl,amsmath,amssymb,showpacs,showkeys,twocolumn]{revtex4}
\usepackage{graphicx}
\usepackage{dcolumn}
\usepackage{bm}
\usepackage{amssymb}
\usepackage{amsmath}
\begin{document}

\title{Strong orientational effect of stretched aerogel on the $^3$He order parameter }

\author{J.~Elbs }
\author{Yu.~M.~Bunkov\/\thanks{e-mail: yuriy.bunkov@grenoble.cnrs.fr}}
\author{E.~Collin}
\author{H.~Godfrin}

\address{Institut N\'eel, CNRS et Universit\'e Joseph Fourier, BP 166, F-38042 Grenoble Cedex 9, France}

\author{G.E. Volovik}
\address{Low Temperature Laboratory, Helsinki University of
Technology, PO BOX 5100, 02015 TKK, Finland\\
and L.D. Landau Institute for Theoretical Physics, 119334
 Moscow, Russia}
\date{\today}

\begin{abstract}

Deformation of aerogel strongly modifies the orientation of the order parameter of superfluid
$^3$He confined in aerogel. We used a radial squeezing of aerogel to keep the orbital angular momentum of the $^3$He Cooper pairs in the plane perpendicular to the magnetic field. 
We did not find strong evidence for a ``polar" phase, with a nodal line  along the equator of the Fermi surface, predicted to occur  at large radial  squeezing. Instead we 
observed $^3$He-A  with a clear experimental evidence of the destruction of the
long-range order by random anisotropy -- the Larkin-Imry-Ma effect.
In $^3$He-B we observed and identified new modes of NMR, which are
impossible to obtain in bulk $^3$He-B. One of these modes is
characterized by a repulsive interaction between magnons, which is
suitable for the magnon Bose-Einstein condensation (BEC).

\end{abstract}



\maketitle

The $p$-wave superfluid $^3$He, which is characterized by an 18-dimensional 
order parameter, is an amazing test system for different aspects of quantum field
theory \cite{Volovik}. Confined in a
silica aerogel, superfluid $^3$He becomes an ideal system for the investigation of the
effect of impurities on a long-range order. In a
first approximation the aerogel reduces the superfluid transition
temperature \cite{Parp,Osh}. Here we report the influence of the aerogel
anisotropy on the $^3$He order parameter orientation. It was demonstrated earlier 
that  a  uniaxial  squeezing of a cylindrical aerogel sample leads to the orientation
of the orbital momentum $\hat{\bf l}$ along the cylinder axis 
\cite{BunkovDeformed}. Here we report  the first  results
of experiments with superfluid $^3$He whose order parameter is
deformed by radially  squeezing the aerogel, which is equivalent  to  uniaxially
stretching  along  the axis of a cylindrical sample. The
particular interest for this kind of deformation is due to the 
prediction of a new phase of superfluid $^3$He -- the
``polar" phase, which has a nodal line in the quasiparticle energy
gap in the plane perpendicular to the stretching  direction \cite{Jap}. In
our experiments no strong evidence of a polar
phase has been found. However, a precursor of a polar phase formation -- orientation of the orbital
vector $\hat{\bf l}$   in the plane of the 
aerogel squeezing in both $^3$He-A and $^3$He-B  -- has been observed. 

This orientational effect allows us to study the influence of the local random anisotropy
of aerogel on the $U(1)$-field of the vector $\hat{\bf l}$, which is kept  in the plane of squeezing. We find
that instead of a polar phase, the Imry-Ma state of superfluid $^3$He-A is formed. 
The quenched random anisotropy of the aerogel strands destroys the long-range
orientational order (LROO) according to the famous Imry-Ma scenario
\cite{ImryMa} (see \cite{Fedorenko,VolovikQFS} and references
therein).  This is the counterpart of the effect of collective
pinning in superconductors predicted by Larkin \cite{Larkin}, in
which  weak impurities destroy the long-range translational  order
of the Abrikosov vortex lattice.

The Larkin-Imry-Ma (LIM) effect has been studied experimentally in  $^3$He-A confined in 
a non-deformed aerogel \cite{BunkovDeformed,Dmitriev}. The aerogel diminishes the
value of Leggett frequency, which leads to  decrease of the
frequency shift by about factor 4 for 98\%
porosity aerogel. This reduction of the Leggett frequency is similar to that
observed in $^3$He-B. For $^3$He-A in a non-deformed aerogel, an additional decrease of
the frequency shift by an order of magnitude has been observed
\cite{Dmitriev}. The additional reduction is an evidence of the disordered LIM state.  
In the non-deformed aerogel, which is globally isotropic and where the vector $\hat{\bf l}$ has a degeneracy on the 
two dimensional unit sphere $S^2$, the  randomization of orientations of $\hat{\bf l}$ by 
the LIM effect leads to an almost complete nullification of the frequency shift \cite{VolovikQFS}.  
It has been found later that the nominal value of the frequency shift is
restored when a sufficiently large axial squeezing is applied
to aerogel \cite{BunkovDeformed}. This occurs because the deformed
aerogel acquires a global anisotropy along the axis of squeezing;
the regular  anisotropy suppresses the Larkin-Imry-Ma (LIM) effect
and induces a homogeneous orientation of $\hat{\bf l}$ in the whole
sample. Uniaxial squeezing of aerogel is a unique tool for
reaching a uniform orientation of the order parameter in $^3$He-A, 
which allows us to study many interesting
effects that are not possible in bulk $^3$He-A
\cite{BunkovDeformed,VolovikQFS}. In addition, the
controllable deformation enables us to study the interplay
between the regular and random anisotropy in the LIM effect.

The radially squeezed aerogel in our experiment adds a new state of  $^3$He-A in aerogel -- the $U(1)$ LIM state.  
The  frequency shift in this state is essentially different from  both  the $S^2$  LIM state 
in a globally isotropic aerogel and the fully oriented state in an axially squeezed aerogel. 
Our observation of the  NMR signatures, which are  in agreement with the $U(1)$ LIM state, 
serves as a final prove of the existence of the LIM effect  in aerogel.

While it is natural that anisotropic $^3$He-A  is influenced by the
random and regular anisotropy of the aerogel strands,  at first
glance no such influence is expected in the case of the isotropic
$^3$He-B. However, superfluid  $^3$He-B has an unusual symmetry
breaking, as shown by Leggett \cite{Leggett}. Although the liquid is
isotropic under combined spin and orbital rotations, symmetry is
broken with respect to the relative rotations of spin and orbital
spaces described by the order parameter matrix $R_{\alpha i}$. As a
consequence, if one introduces  a nonzero spin density ${\bf S}$
by applying a magnetic field and thus creates anisotropy in the spin
space, one automatically creates anisotropy in the orbital space.
The gap in the quasiparticle spectrum becomes anisotropic, it is
smaller along the axis $\hat{\bf l}$, which is connected to the
direction of spin $\hat{\bf s}$ by the  order parameter matrix:
$\hat l_i=R_{\alpha i} \hat s_\alpha$. The magnitude of the gap
distortion is determined by the Larmor frequency and increases with
increasing field: $\Delta_\perp^2-\Delta_\parallel^2\sim\omega_L^2$
(see Ref. \cite{Pickett1} and references therein).

Thus in an applied magnetic field the aerogel strands do influence
the orientation of the order parameter. However, this orientational
effect is by a factor $\omega_L^2/\Delta_\perp^2$ weaker than in the
gapless $^3$He-A, where $\Delta_\parallel=0$. Correspondingly, the
Larkin-Imry-Ma length, at which the orientational order is destroyed
by random anisotropy of aerogel, is $L_{\rm LIM}\sim 1~\mu$m in
$^3$He-A  \cite{VolovikQFS} and  by a factor
$\Delta_\perp^4/\omega_L^4$ larger in $^3$He-B.  In typical NMR
experiments in $^3$He-B  with moderate $\omega_L\sim 1~$MHz, $L_{\rm
LIM}$ essentially exceeds the sample size, and thus the random
anisotropy has no effect on the orientation of $\hat{\bf l}$.

In  experiments with uniaxial squeezing of aerogel along the
magnetic field ${\bf H}$ the orientation of $\hat{\bf l}$ in $^3$He-B is fixed
along the field,  $\hat{\bf l}\parallel  {\bf H}$ \cite{BunkovDeformed}. Experiments with a
radially squeezed (uniaxially stretched) aerogel, described in this
article, revealed a strong reorientation of $\hat{\bf l}$
perpendicular to magnetic field, $\hat{\bf l}\perp  {\bf H}$. This effect can be explained and
estimated in a simplified model of aerogel deformation (see for
details \cite{VolovikQFS}), in which aerogel is considered as a
percolating cluster of random cylinders of diameter $\delta\sim 3$
nm  and the length $\xi_{\rm a}\sim $ 20 nm corresponding to the
diameter of and the distance between the silica strands. Applying
the theory of Rainer and Vuorio \cite{RainerVuorio} for a
microscopic cylindrical body with a diameter $\delta\ll \xi_0$, where $\xi_0$ is the
superfluid coherence length, one obtains the estimate for the energy
density  of the interaction between $\hat{\bf l}$  and the global
uniaxial deformation along the axis $\hat{\bf \nu}$ both in $^3$He-B
in magnetic field and in $^3$He-A:
\begin{equation}
E_{\rm an}= C  (\hat{\bf l}\cdot \hat{\bf\nu})^2 ~~,~~ C\sim N_F
(\Delta_\perp^2-\Delta_\parallel^2) \frac{\Delta l}{l}
\frac{\xi_0\delta}{\xi_{\rm a}^2} ~. \label{RegularAnisotropyEnergy}
\end{equation}
Here $N_F$ is the density of states in normal $^3$He, and $\Delta
l/l$ is the relative change  of the aerogel sample length. For a
typical sample of cylindrical shape, squeezing ($\Delta l<0$)
produces an easy axis anisotropy for $\hat{\bf l}$, as was observed
in Ref. \cite{BunkovDeformed}  for $^3$He-A, while stretching
($\Delta l>0$) should give an easy plane. In $^3$He-B this energy is
rather small, but it may compete  with another energy which is
responsible for the orientation of the order parameter
 -- the tiny  spin-orbit (dipole-dipole) coupling between
$\hat{\bf l}$ and $\hat{\bf s}$. The dipole interaction $E_{\rm D}$
is characterized by the Leggett frequency $\Omega_B$ which enters
the NMR frequency shift from the Larmor value in $^3$He-B caused by
the dipole interaction and  is on the order of $E_{\rm D} \sim N_F
\Omega_B^2 $. The global anisotropy of aerogel prevails if
\begin{equation}
\frac{|\Delta l |}{l} > \frac{\xi_{\rm a}^2}{\xi_0\delta}
\frac{\Omega_B^2}{\omega_L^2} ~. \label{Criterium}
\end{equation}
With $\xi_0\sim \xi_{\rm a}$, $\delta \sim 0.1\xi_{\rm a}$,
$\omega_L\sim 1~$MHz and $\Omega_B\sim 100~$kHz, one obtains that
squeezing or stretching the aerogel sample by about 10\% may lead to
the global orientation of the order parameter in $^3$He-B, as was
found in experiments described below.

\begin{figure}[htt]
 \includegraphics[width=0.35\textwidth]{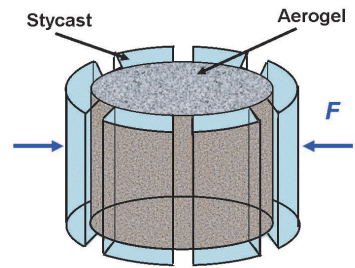}
 \caption{Formation of the global easy plane anisotropy in the aerogel sample by
 radial squeezing.}
 \label{Sample}
\end{figure}

A global anisotropy of aerogel can be achieved in two different
ways: it can be introduced during the aerogel synthesis with rapid
supercritical extraction or by mechanical squeezing of a homogeneous
aerogel (for details see the recent paper \cite{Halperin}). Since we
already had 98\% aerogel samples, prepared by N. Mulders, we
 used  the mechanical squeezing method. To deform the aerogel
significantly, we placed the cylindrical aerogel sample of
diameter 5 mm and 12 mm length in a
 tube of 0.5 mm thickness, and internal diameter 5 mm
with 6 longitudinal slots of 0.5 mm thickness. (see Fig.
\ref{Sample}). Then the tube with the sample was pressed into
another  tube with inner diameter of 5 mm and conical entrance. As a
results the slots were closed, and the inner diameter of the cell
together with the diameter of aerogel became only 4 mm. By this
method we have squeezed  the aerogel sample in a plane,
perpendicular to its axis (i.e. $\Delta l>0$ in
Eq.(\ref{RegularAnisotropyEnergy})) by about 20\%.
 Then we glued the tube at the bottom of the cell and
mounted it in a demagnetization refrigerator with a magnetic field
along the cell axis $ {\bf H}/H=\hat{\bf\nu}\equiv \hat{\bf z}$,
and filled the cell with superfluid $^3$He.

In order to avoid signals from paramagnetic solid $^3$He on the
surface of aerogel strands, we added some amount of $^4$He, which
covered the strands.  The amplitude of the RF field was calibrated
in normal $^3$He: after applying five periods of RF oscillations the
magnetization is deflected by 90$^\circ$. The homogeneity of the RF
field was about 10\% within 8 mm of the sample.

\begin{figure}[htt]
 \includegraphics[width=0.45\textwidth]{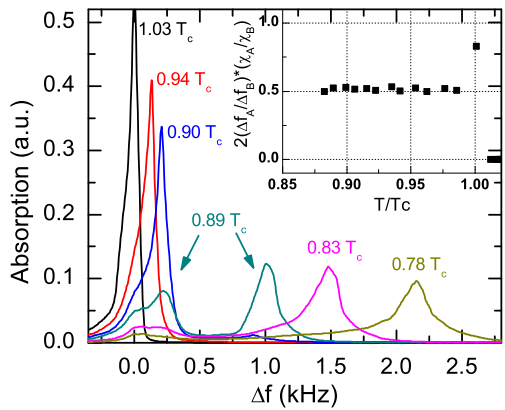}
 \caption{The NMR frequency shift for normal $^3$He (1.83 mK),
 $^3$He-A (1.47 - 1.39 mK) and $^3$He-B (1.39 - 1.22 mK) measured at 25 bar and $\omega/2\pi=1177.4$ kHz.
The temperature dependence of the factor $F$ in Eqs. (\ref{RatioDelta}) and (\ref{APhase}) is shown in inset.}
 \label{Sample}
\end{figure}

First we have measured by cw NMR the frequency shift as a
function of temperature. The raw data at cooling  are
shown in Fig.2 for 25 bar. A clear coexistence of A and B phases is seen at $T=1.39$ mK. The important information about the
structure of superfluid states can be extracted from the ratio of
the corresponding Leggett frequencies $\Omega_B^2/\Omega_A^2$
\cite{Leggett}:
\begin{equation}
\frac{\Omega_B^2}{\Omega_A^2}  =
\frac{5}{2}\frac{\chi_A}{\chi_B}\frac{\Delta_B^2}{\Delta_A^2}~.
 \label{B/A}
\end{equation}
This equation with the ratio of gap parameters $\Delta_B/\Delta_A$ close to 1,  was confirmed for bulk $^3$He in the plate geometry \cite{Ahonen} and in other experiments
(see p.103 in \cite{VollhardtWolfle}). Since the
maximum of the NMR frequency shift in $^3$He-A and $^3$He-B for
$\hat{\bf l}$ oriented perpendicular to $\hat{\bf H}$ is:
\begin{equation}
 \Delta \omega_A  = \frac{ \Omega_A^2}{2 \omega_L}~~, \, \, \,\Delta \omega_B  = \frac{4}{5} \frac{ \Omega_B^2}{
 2\omega_L}~,
 \label{DeltaAB}
\end{equation}
the ratio of frequency shifts should be:
\begin{equation}
 \frac{\Delta \omega_A}{\Delta \omega_B} = F\,  \frac{\chi_B}{2 \chi_A},
 \label{RatioDelta}
\end{equation}
where the factor F is incorporated to characterize the deviation
of the experimental data from the theoretical value $F=1$. From our data we
directly obtain the factor F, which is about 0.5 in the radially squeezed
aerogel (see inset in Fig.2; we have measured ${\Delta \omega_B}$ and $\chi_B$
in the considered range of temperatures at warming). How to explain such a
small F factor?

The aerogel sample exhibiting about  20\%  of radial shrinkage is
expected to orient  $\hat{\bf l}\perp {\bf H}$ in both phases of
$^3$He. As distinct from the axial squeezing  which completely removes the 
$S^2$ degeneracy of $\hat{\bf l}$ in the non-deformed aerogel, in the radially squeezed aerogel  the continuous $U(1)$ degeneracy remains, since the orientation of $\hat{\bf l}$ in the
plane perpendicular to ${\bf H}$ is not fixed.  Because of the remaining 
degeneracy  the LIM effect is still operating, but its realization
may differ from the non-Abelian $S^2$ case  \cite{VolovikQFS}. For the
Abelian group,  the quasi-long-range order (QLRO) with  a power-law
decay of correlators is possible, as was discussed for the
translational $U(1)\times U(1)$ group of a vortex lattice in
superconductors \cite{Giamarchi}. The LIM effect is not relevant for
$^3$He-B, where $L_{\rm LIM}$ exceeds the sample size, but is
important for $^3$He-A, where $L_{\rm LIM}$ is smaller than the
dipole length $\xi_D$, characterizing  spin-orbit interaction
between the orbital vector $\hat{\bf l}$ and the spin-nematic vector
$\hat{\bf d}$ \cite{VolovikQFS}. The general expression for the frequency shift in  $^3$He-A
is:
\begin{equation}
 \omega-\omega_L= F\frac{\Omega_A^2}{2 \omega_L} ~~,~~F=
 (\hat{\bf l}\cdot \hat{\bf d})^2-(\hat{\bf l}\cdot \hat{\bf h})^2~,~ \hat{\bf h}=\frac {\bf H}{H}~.
 \label{APhase}
\end{equation}
Since ${\bf H}$  is applied along the sample axis, both $\hat{\bf
l}$ and  $\hat{\bf d}$ are kept in the plane: $\hat{\bf d}\perp {\bf
H}$ due to orientation by magnetic field and $\hat{\bf l}\perp {\bf
H}$ due to radial squeezing.  In the uniform texture, one has
$\hat{\bf d}\parallel\hat{\bf l}$ due to spin-orbit  interaction,
and the factor $F =1$. The factor $F=1/2$ found in our experiment is readily
explained by the $U(1)$ LIM effect. In the LIM state, the $\hat{\bf
l}$-vector is disordered, while $\hat{\bf d}$ remains uniform at LIM
scale because $L_{\rm LIM}<\xi_D$. For the $S^2$ LIM effect in the
non-deformed aerogel, where $\hat{\bf l}$ is allowed to have all
orientations, the factor $F$ is   small since $F=\left<(\hat{\bf
l}\cdot \hat{\bf d})^2\right>- \left<(\hat{\bf l}\cdot \hat{\bf
h})^2\right>\approx \frac{1}{3} - \frac{1}{3} \approx 0$. This leads
to the small frequency shift observed in the non-deformed aerogel
\cite{Dmitriev}. For the $U(1)$ LIM effect. i.e. for planar orientation of $\hat{\bf l}$ one
obtains  $F=\left< (\hat{\bf l}\cdot \hat{\bf d})^2\right>
=\frac{1}{2}$.

Now we demonstrate that in $^3$He-B the radial deformation also
orients the orbital momentum  in the plane of squeezing. For that
let us compare the NMR frequency shifts for two orientations  of
$\hat{\bf l}$ in $^3$He-B. If  $\hat{\bf l}\parallel {\bf H}$, there
are two NMR modes characterized by the dependence of the frequency
shift  on the tipping angle of magnetization $\beta$ --
Brinkman-Smith (BS) mode for $ \beta < 104^\circ$  \cite{BS} and
Osheroff-Corruccini mode (OC) for  $\beta > 104^\circ$ \cite{OC}:
 \begin{eqnarray}
 \omega-\omega_L= 0,~\beta < 104^\circ~({\rm BS})~,
\label{BS}
\\
 \omega-\omega_L= -\frac{16}{15}\frac{\Omega_B^2}{\omega_L}\left( \cos\beta+\frac{1}{4}\right)
,~\beta > 104^\circ~({\rm OC})~.
\label{OC}
\end{eqnarray}
For the transverse orientation $\hat{\bf l}\perp {\bf H}$ one has
\cite{BV2,DV}:
\begin{eqnarray}
 \omega-\omega_L= \frac{\Omega_B^2}{2 \omega_L}\left(\cos\beta - \frac{1}{5}\right),~\beta < 90^\circ~({\rm
 mode~I})~,
  ~\label{SpinWave2}
\\
 \omega-\omega_L= - \frac{\Omega_B^2}{10 \omega_L}(1+ \cos\beta),~\beta > 90^\circ ~({\rm
 mode~II})~.
 ~\label{SpinWave3}
\end{eqnarray}

\begin{figure}[htt]
 \includegraphics[width=0.4\textwidth]{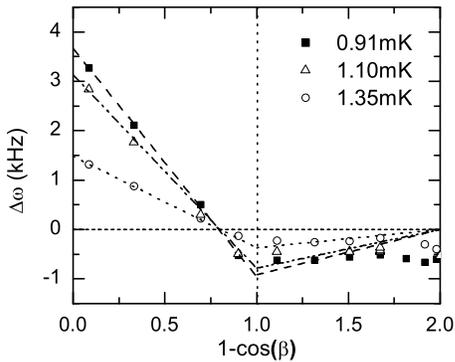}
 \caption{The frequency shift $\omega-\omega_L$ at the beginning of the induction signal decay
 after a pulse which deflects the magnetization by an angle $\beta$.
 The lines are theoretical values  of  the frequency shift in Eqs. (\ref{SpinWave2}) and (\ref{SpinWave3}) at a given temperature.}
 \label{frequency}
\end{figure}

We have measured the NMR frequency of the induction signal after 
deflection of magnetization by different angles $\beta$ in
$^3$He-B in the radially squeezed aerogel. The experimental results
in Fig.(\ref{frequency}) obtained at different temperatures  are in
good agreement with Eqs.~(\ref{SpinWave2}) and (\ref{SpinWave3})
confirming that  $\hat{\bf l}$ is kept in the plane of squeezing. The
deviation from the theoretical curves for  angles above
140$^\circ$ may have the same origin as a similar deviation observed in  bulk  $^3$He-B
for large $\beta$ angles \cite{OsheroffB}, where the spin dynamics becomes more complicated.


In conclusion, by immersing superfluid $^3$He in a radially squeezed aerogel, we
obtained a new stable orientation of the orbital momentum $\hat{\bf
l}$ in both phases, $^3$He-A and $^3$He-B. For this orientation two
new NMR modes of precession have been observed in $^3$He-B. The mode
I with a magnetization deflection $\beta < 90^\circ$   is
unstable because in the magnon BEC presentation of the coherent spin 
precession the interaction between magnons  is attractive. 
On the contrary, for the mode II with  $\beta >
90^\circ$, the free precession is stable because of the repulsive magnon interaction. 
This means that the mode II may form the BEC state of magnons \cite{BV2}. 
The observed long induction decay
after a $120^\circ$ pulse, which will be discussed elsewhere, is in
favor of the identification of the mode II with the HPD2 \cite{BV}
-- the new phase-coherent self-sustained state of precession which
must be added to the known BEC states of magnons: HPD and ``Q-ball''
\cite{Q}.

 In $^3$He-A in the radially squeezed aerogel, we found experimental evidence
for the Larkin-Imry-Ma state, in which the long range order of the
orbital vector $\hat{\bf l}$ is destroyed by the random anisotropy of
the aerogel strands. While in the non-deformed aerogel the
Larkin-Imry-Ma state is described by  the non-Abelian $SO(3)$ group,  in the
radially squeezed aerogel  the $U(1)$ Larkin-Imry-Ma state takes place.
We found that the NMR signatures of these two disordered states are
essentially different.

We are grateful to M. Krusius  for illuminating discussions and N.
Mulders for providing the aerogel samples. This work was done in the framework of the Large Scale Installation Program
ULTI of the European Union (contract RITA-CT-2003-505313), a
collaboration between CNRS and the Russian Academy of Sciences
(project N16569), and was supported in part by the Russian
Foundation for Basic Research (grant 06-02-16002-a).

\end{document}